\documentclass[8pt]{revtex4}
\usepackage{amssymb}
\usepackage{latexsym}
\usepackage{epsfig}
\begin{document}

\title{Expansion and contraction of accretion disks of a rotating thermal BIon in a  Rindler space-time }

\author{ Aroonkumar Beesham$^{1}$\footnote{beeshama@unizulu.ac.za}, Alireza Sepehri$^{1,2}$\footnote{alireza.sepehri3@gmail.com}}
\affiliation{$^{1}$Department of Mathematical Sciences, University of Zululand, Private Bag X1001, Kwa-Dlangezwa 3886, South Africa\\$^{2}$Research Institute for Astronomy and Astrophysics of
Maragha (RIAAM), P.O. Box 55134-441, Maragha, Iran }

\begin{abstract}
In this paper, we consider evolutions of accretion disks of a rotating BIon in a Rindler space-time. This space-time is emerged by acceleration of disks in BIon. A BIon is constructed from two accretion disks that are connected by a wormhole. We will show that in a rotating BIon, by increasing rotating velocity, area of one accretion disk grows, while, area of other disk decreases. Also, we consider about four types of accreion disks which are produced in Rindler space-time.
\vspace{5mm}\noindent\\
PACS numbers: 98.80.-k, 04.50.Gh, 11.25.Yb, 98.80.Qc\vspace{0.8mm}\newline Keywords: accretion disk, Rotation, Black hole, Temperature
\end{abstract}

\maketitle
\section{Introduction}

Recently, accretion disks around wormhole or black holes have been investigated in some researches \cite{w1,w2,w3}. For example, in one work, thick disk accretions in Kerr space-time with arbitrary spin parameters have been investigated \cite{w3}.  In another investigation, authors have proposed a code based on a ray-tracing approach and capable of computing some basic properties of thin accretion disks in space-times with deviations from the Kerr background. The code has been used to fit current and future X-ray data of stellar-mass black hole candidates and constrain possible deviations from the Kerr geometry in the spin parameter-deformation parameter plane \cite{w2}. Also, in one research,   authors have discussed that the line produced in accretion disks around non-rotating or very slow-rotating wormholes is relatively similar to the one expected around Kerr black holes with mid or high value of spin parameter. They have shown that current observations are still marginally compatible with the possibility that the supermassive black hole candidates in galactic nuclei are sources of these disks\cite{w1}.   Now,the question arises that what is the origin of these disks. In some researches, it has been shown that a accretion disk could be a part of BIon \cite{w4,w5}. Motivated by these researches, we suggest a mathematical model for a rotating accretion disk.  We argue that each accretion disk connects to another disk and builds a BIon. In a rotating BIon, by increasing rotating velocity, one disk expands and another contracts. Previously, various types of BIons have been considered \cite{w4,w5}. Specially, the effect of acceleration on evolutions of BIon has been studied \cite{w4}. In this research, we use of the relation between temperature and acceleration and re-consider a rotating BIon with two accretion disks in thermal system.

The outline of the paper is as follows: In section II, we will consider the  quantum expanding accretion disk in one end of a rotating BIon . In section III, we  will analyse the contraction of accretion disk in a rotating BIon.  The last section is devoted to summary and conclusion.

\section{ Expansion of one accreion disk of a rotating BIon in a Rindler space-time}\label{o1}

To consider the rotating accretion disk, we specialize to an embedding of the disk world volume in
Minkowski space-time with metric \cite{w4,w5};

\begin{eqnarray}
&& ds^{2}=-dt^{2} + dr^{2} + r^{2}\Big(d\theta^{2} + sin^{2}\theta d\phi^{2}\Big) + \sum_{i=1}^{6}dx_{i}^{2}
\label{a1}
\end{eqnarray}

without background fluxes. We assume that disks are rotated with the acceleration. In this case, the relation between  the world volume coordinates of the rotating accretion disk ($\tau, \sigma $) and the coordinates of
Minkowski space-time ($t, r$) are \cite{w4};

\begin{eqnarray}
&& at= e^{a\sigma} \sinh(a\tau) \quad ar=e^{a\sigma} \cosh(a\tau) \quad \text{In Region I} \nonumber\\&& at= - e^{-a\sigma} \sinh(a\tau) \quad ar =  e^{-a\sigma} \cosh(a\tau) \quad \text{In Region II}
\label{a2}
\end{eqnarray}

where $a$ is the acceleration of rotating disk. The acceleration leads to the emergence of two new regions in a Rindler space-time. In each region, we have a  accretion disk. The behaviour of the accretion disk in region I is reverse to the accretion disk in region II.

\begin{figure*}[thbp]
	\begin{center}
		\begin{tabular}{rl}
			\includegraphics[width=14cm]{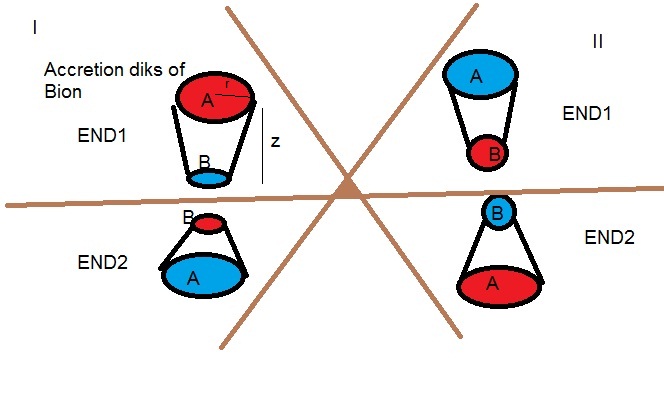}
		\end{tabular}
	\end{center}
	\caption{  Accretion disks of BIon in the Rindler space-time}
\end{figure*}

\begin{figure*}[thbp]
	\begin{center}
		\begin{tabular}{rl}
			\includegraphics[width=14cm]{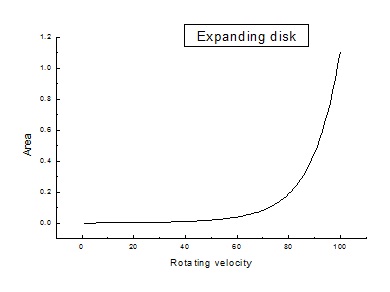}
		\end{tabular}
	\end{center}
	\caption{  Expanding disk. Area in terms of rotating velocity}
\end{figure*}

We can suppose that the coordinate along the separation distance between disks ($ x^{4} = z$) depends on the $r =\pm\frac{1}{a} e^{\pm a\sigma} \cosh(a\tau)$ and by using equations (\ref{a2}), rewrite equations (\ref{a1}) as \cite{w4};

\begin{eqnarray}
&& ds^{2}_{I}=-dt^{2} + \Big(1 + (\frac{dz}{dr})^{2} \Big) dr^{2} + r^{2}\Big(d\theta^{2} + sin^{2}\theta d\phi^{2}\Big) + \sum_{i=1}^{5}dx_{i}^{2} = \nonumber\\&&  \Big(e^{2a\sigma} + \frac{1}{\sinh^{2}(a\tau)}(\frac{dz}{d\tau})^{2} \Big)d\tau^{2} -  \Big(e^{2a\sigma}+ \frac{1}{\cosh^{2}(a\tau)}(\frac{dz}{d\sigma})^{2} \Big) d\sigma^{2} + \nonumber\\&& \frac{1}{\sinh(a\tau)\cosh(a\tau)}(\frac{dz}{d\tau }\frac{dz}{d\sigma})d\tau d\sigma + \Big(\frac{1}{a}e^{a\sigma} \cosh(a\tau)\Big)^{2}\Big(d\theta^{2} + sin^{2}\theta d\phi^{2}\Big)  + \sum_{i=1}^{5}dx_{i}^{2}
\label{a5}
\end{eqnarray}

\begin{eqnarray}
&& ds^{2}_{II}=-dt^{2} + \Big(1 + (\frac{dz}{dr})^{2} \Big) dr^{2} + r^{2}\Big(d\theta^{2} + sin^{2}\theta d\phi^{2}\Big) + \sum_{i=1}^{5}dx_{i}^{2} = \nonumber\\&&  \Big(e^{-2a\sigma} + \frac{1}{\sinh^{2}(a\tau)}(\frac{dz}{d\tau})^{2} \Big)d\tau^{2} -  \Big(e^{-2a\sigma}+ \frac{1}{\cosh^{2}(a\tau)}(\frac{dz}{d\sigma})^{2} \Big) d\sigma^{2} -\nonumber\\&& \frac{1}{\sinh(a\tau)\cosh(a\tau)}(\frac{dz}{d\tau }\frac{dz}{d\sigma})d\tau d\sigma + \Big(\frac{1}{a}e^{-a\sigma} \cosh(a\tau)\Big)^{2}\Big(d\theta^{2} + sin^{2}\theta d\phi^{2}\Big)  + \sum_{i=1}^{5}dx_{i}^{2}
\label{a6}
\end{eqnarray}

These equations show that the rotation of disks leads to the emergence of a new Rindler space-time. Now, we can replace acceleration with it's equivalent temperature.
Previously,it has been shown that temperature has the below relation with acceleration \cite{w4,w5}:

\begin{eqnarray}
&& v=a\tau \nonumber\\&&\rightarrow T=\frac{T_{0}}{\sqrt{1-\frac{v^{2}}{c^{2}} }}=\frac{T_{0}}{\sqrt{1-\frac{[a\tau]^{2}}{c^{2}}} }\nonumber\\&& v=a\tau= c \sqrt{1-\frac{T_{0}^{2}}{T^{2}}}
\label{a22}
\end{eqnarray}
where $T$ is temperature of the BIon and $T_{0}$ is the critical temperature relating to the colliding point of the branes. However this elation is questionable. Based on this relation, the superconductivity phenomena depends on the system velocity!! You can move a system with special velocities to reduce its temperature to the less than of its critical temperature and then the system shows superconductivity by itself!! In fact, it means that a physical phenomena (superconductivity) depends on the system velocity, a result in direct conflict with the relativity law claiming that the physical laws are independent of the observer velocity. This relativistic relation for temperature is not a true relation, and in fact, the temperature's relation depends on the thermocouple apparatus used. A true thermocouple rejects this definition of temperature (For example, see \cite{E1,E2,E3}). Thus, to obtain true relation between temperature and acceleration, we use of concepts of BIon:

\begin{eqnarray}
&& d M_{I-A/B} = T_{I-A/B} d S_{I-A/B} \rightarrow T_{I-A/B}=\frac{d M_{I-A/B}}{d S_{I-A/B}}
\label{relation1}
\end{eqnarray}

Previously, thermodynamical parameters have been obtained in \cite{q12}:

\begin{eqnarray}
&& d M_{I-A} = \frac{d M_{I-A}}{dz_{I-A}}dz_{I-A} \nonumber\\&&  d z_{I-A} = dz_{II-B}\simeq \Big(e^{-4a\sigma}\sinh^{2}(a\tau)\cosh^{2}(a\tau)\Big) \times \nonumber\\&& \Big(\frac{F_{DBI,I,A}(\tau,\sigma)\Big(\frac{F_{DBI,I,A}(\tau,\sigma)}{F_{DBI,I,A}(\tau,\sigma_{0})}-e^{-4a(\sigma-\sigma_{0})}\frac{\cosh^{2}(a\tau_{0})}{\cosh^{2}(a\tau)}\Big)^{-\frac{1}{2}}}{F_{DBI,I,A}(\tau_{0},\sigma)\Big(\frac{F_{DBI,I,A}(\tau_{0},\sigma)}{F_{DBI,I,A}(\tau_{0},\sigma_{0})}-e^{-4a(\sigma-\sigma_{0})}\frac{\cosh^{2}(a\tau_{0})}{\cosh^{2}(a\tau)}\Big)^{-\frac{1}{2}}}-\frac{\sinh^{2}(a\tau_{0})}{\sinh^{2}(a\tau)}\Big)^{-\frac{1}{2}}\nonumber\\&& \frac{d M_{I-A}}{dz}=  \frac{d M_{II-B}}{dz}=\frac{4 T_{D3}^{2}}{\pi T_{0,I-A}^{4}}  \frac{F_{DBI,I,A}(\sigma,\tau)\Big(\frac{1}{a}e^{a\sigma} \cosh(a\tau)\Big)^{2}\Big(\sinh^{2}(a\tau)+cosh^{2}(a\tau)\Big)}{\sqrt{F_{DBI,I,A}^{2}(\sigma,\tau)-F_{DBI,I,A}^{2}(\sigma_{o},\tau)}}\times \nonumber\\&& \frac{4 \cosh^{2}\alpha_{I-A} + 1}{\cosh^{4}\alpha_{I-A}}\nonumber\\&& d S_{I-A}=d S_{II-B}=\frac{4 T_{D3}^{2}}{\pi T_{0,I-A}^{5}}    \frac{F_{DBI,I,A}(\sigma,\tau)\Big(\frac{1}{a}e^{a\sigma} \cosh(a\tau)\Big)^{2}\Big(\sinh^{2}(a\tau)+cosh^{2}(a\tau)\Big)}{\sqrt{F_{DBI,I,A}^{2}(\sigma,\tau)-F_{DBI,I,A}^{2}(\sigma_{o},\tau)}}\times \nonumber\\&& \frac{4 }{\cosh^{4}\alpha_{I-A}}
\label{relation2}
\end{eqnarray}

and

\begin{eqnarray}
&&  d M_{I-B} = \frac{d M_{I-B}}{dz_{I-B}}dz_{I-B} \nonumber\\&& d z_{I-B} = d z_{II-A}\simeq \Big(e^{4a\sigma}\sinh^{2}(a\tau)\cosh^{2}(a\tau)\Big) \times \nonumber\\&& \Big(\frac{F_{DBI,II,A}(\tau,\sigma)\Big(\frac{F_{DBI,II,A}(\tau,\sigma)}{F_{DBI,II,A}(\tau,\sigma_{0})}-e^{4a(\sigma-\sigma_{0})}\frac{\cosh^{2}(a\tau_{0})}{\cosh^{2}(a\tau)}\Big)^{-\frac{1}{2}}}{F_{DBI,II,A}(\tau_{0},\sigma)\Big(\frac{F_{DBI,II,A}(\tau_{0},\sigma)}{F_{DBI,II,A}(\tau_{0},\sigma_{0})}-e^{4a(\sigma-\sigma_{0})}\frac{\cosh^{2}(a\tau_{0})}{\cosh^{2}(a\tau)}\Big)^{-\frac{1}{2}}}-\frac{\sinh^{2}(a\tau_{0})}{\sinh^{2}(a\tau)}\Big)^{-\frac{1}{2}}   \nonumber\\&&   \frac{d M_{I-B}}{dz}= \frac{d M_{II-A}}{dz}=\frac{4 T_{D3}^{2}}{\pi T_{0,II-A}^{4}}    \frac{F_{DBI,II,A}(\sigma,\tau)\Big(\frac{1}{a}e^{-a\sigma} \cosh(a\tau)\Big)^{2}\Big(\sinh^{2}(a\tau)+cosh^{2}(a\tau)\Big)}{\sqrt{F_{DBI,II,A}^{2}(\sigma,\tau)-F_{DBI,II,A}^{2}(\sigma_{o},\tau)}}\times \nonumber\\&& \frac{4 \cosh^{2}\alpha_{II-A} + 1}{\cosh^{4}\alpha_{II-A}}\nonumber\\&& d S_{II-A}= d S_{I-B}=\frac{4 T_{D3}^{2}}{\pi T_{0,II-A}^{5}}  \frac{F_{DBI,II,A}(\sigma,\tau)\Big(\frac{1}{a}e^{-a\sigma} \cosh(a\tau)\Big)^{2}\Big(\sinh^{2}(a\tau)+cosh^{2}(a\tau)\Big)}{\sqrt{F_{DBI,II,A}^{2}(\sigma,\tau)-F_{DBI,II,A}^{2}(\sigma_{o},\tau)}}\times \nonumber\\&& \frac{4 }{\cosh^{4}\alpha_{II-A}}\label{relation3}
\end{eqnarray}

using relation (\ref{relation2} and \ref{relation3}) in relation (\ref{relation1}),we can obtain explicit form of temperature in an accelerating BIon:

\begin{eqnarray}
&&  T_{I-A}= T_{0,I-A} \Big(4 \cosh^{2}\alpha_{I-A} + 1\Big)\times \nonumber\\&&\Big(e^{-4a\sigma}\sinh^{2}(a\tau)\cosh^{2}(a\tau)\Big) \times \nonumber\\&& \Big(\frac{F_{DBI,I,A}(\tau,\sigma)\Big(\frac{F_{DBI,I,A}(\tau,\sigma)}{F_{DBI,I,A}(\tau,\sigma_{0})}-e^{-4a(\sigma-\sigma_{0})}\frac{\cosh^{2}(a\tau_{0})}{\cosh^{2}(a\tau)}\Big)^{-\frac{1}{2}}}{F_{DBI,I,A}(\tau_{0},\sigma)\Big(\frac{F_{DBI,I,A}(\tau_{0},\sigma)}{F_{DBI,I,A}(\tau_{0},\sigma_{0})}-e^{-4a(\sigma-\sigma_{0})}\frac{\cosh^{2}(a\tau_{0})}{\cosh^{2}(a\tau)}\Big)^{-\frac{1}{2}}}-\frac{\sinh^{2}(a\tau_{0})}{\sinh^{2}(a\tau)}\Big)^{\frac{1}{2}}\label{relation5}
\end{eqnarray}

and

\begin{eqnarray}
&& T_{I-B}= T_{0,I-B} \Big(4 \cosh^{2}\alpha_{I-B} + 1\Big)\times \nonumber\\&&\Big(e^{4a\sigma}\sinh^{2}(a\tau)\cosh^{2}(a\tau)\Big) \times \nonumber\\&& \Big(\frac{F_{DBI,I,B}(\tau,\sigma)\Big(\frac{F_{DBI,I,B}(\tau,\sigma)}{F_{DBI,I,B}(\tau,\sigma_{0})}-e^{4a(\sigma-\sigma_{0})}\frac{\cosh^{2}(a\tau_{0})}{\cosh^{2}(a\tau)}\Big)^{-\frac{1}{2}}}{F_{DBI,I,B}(\tau_{0},\sigma)\Big(\frac{F_{DBI,I,B}(\tau_{0},\sigma)}{F_{DBI,I,B}(\tau_{0},\sigma_{0})}-e^{4a(\sigma-\sigma_{0})}\frac{\cosh^{2}(a\tau_{0})}{\cosh^{2}(a\tau)}\Big)^{-\frac{1}{2}}}-\frac{\sinh^{2}(a\tau_{0})}{\sinh^{2}(a\tau)}\Big)^{\frac{1}{2}}
\label{relation6}
\end{eqnarray}

Above equations show the explicit relation between temperatures and acceleration in BIon.However to obtain the relation between temperature and rotating velocity, we should take a derivation of above equations, put ( $\omega= \frac{d\sigma}{dt}$) and obtain below relation:

\begin{eqnarray}
&& a \sim 2\pi T=  \frac{T_{0}ln^{-1}[1-\frac{\omega^{2}}{\omega_{0}^{2}}]}{[1-\frac{\omega^{2}}{\omega_{0}^{2}}]^{\frac{1}{4}}}
\label{thermal1}
\end{eqnarray}

where $T_{0}$ is temperature of non-rotating accretion disk and $\omega$ is the rotating velocity.

Substituting equation (\ref{thermal1}) in equations (\ref{a5}, \ref{a6}),we obtain:

\begin{eqnarray}
&& ds^{2}_{I}= \nonumber\\&&  \Big(e^{2\frac{T_{0}ln^{-1}[1-\frac{\omega^{2}}{\omega_{0}^{2}}]}{[1-\frac{\omega^{2}}{\omega_{0}^{2}}]^{\frac{1}{4}}}\sigma} + \frac{1}{\sinh^{2}(\frac{T_{0}ln^{-1}[1-\frac{\omega^{2}}{\omega_{0}^{2}}]}{[1-\frac{\omega^{2}}{\omega_{0}^{2}}]^{\frac{1}{4}}}\tau)}(\frac{dz}{d\tau})^{2} \Big)d\tau^{2} - \nonumber\\&& \Big(e^{2\frac{T_{0}ln^{-1}[1-\frac{\omega^{2}}{\omega_{0}^{2}}]}{[1-\frac{\omega^{2}}{\omega_{0}^{2}}]^{\frac{1}{4}}}\sigma}+ \frac{1}{\cosh^{2}(\frac{T_{0}ln^{-1}[1-\frac{\omega^{2}}{\omega_{0}^{2}}]}{[1-\frac{\omega^{2}}{\omega_{0}^{2}}]^{\frac{1}{4}}}\tau)}(\frac{dz}{d\sigma})^{2} \Big) d\sigma^{2} + \nonumber\\&& \frac{1}{\sinh(\frac{T_{0}ln^{-1}[1-\frac{\omega^{2}}{\omega_{0}^{2}}]}{[1-\frac{\omega^{2}}{\omega_{0}^{2}}]^{\frac{1}{4}}}\tau)\cosh(\frac{T_{0}ln^{-1}[1-\frac{\omega^{2}}{\omega_{0}^{2}}]}{[1-\frac{\omega^{2}}{\omega_{0}^{2}}]^{\frac{1}{4}}}\tau)}(\frac{dz}{d\tau }\frac{dz}{d\sigma})d\tau d\sigma +\nonumber\\&& \Big(\frac{1}{\frac{T_{0}ln^{-1}[1-\frac{\omega^{2}}{\omega_{0}^{2}}]}{[1-\frac{\omega^{2}}{\omega_{0}^{2}}]^{\frac{1}{4}}}}e^{\frac{T_{0}ln^{-1}[1-\frac{\omega^{2}}{\omega_{0}^{2}}]}{[1-\frac{\omega^{2}}{\omega_{0}^{2}}]^{\frac{1}{4}}}\sigma} \cosh(\frac{T_{0}ln^{-1}[1-\frac{\omega^{2}}{\omega_{0}^{2}}]}{[1-\frac{\omega^{2}}{\omega_{0}^{2}}]^{\frac{1}{4}}}\tau)\Big)^{2}\times \nonumber\\&&\Big(d\theta^{2} + sin^{2}\theta d\phi^{2}\Big)  + \nonumber\\&& \sum_{i=1}^{5}dx_{i}^{2}
\label{thermal2}
\end{eqnarray}

\begin{eqnarray}
&& ds^{2}_{II}= \nonumber\\&&  \Big(e^{-2\frac{T_{0}ln^{-1}[1-\frac{\omega^{2}}{\omega_{0}^{2}}]}{[1-\frac{\omega^{2}}{\omega_{0}^{2}}]^{\frac{1}{4}}}\sigma} + \frac{1}{\sinh^{2}(\frac{T_{0}ln^{-1}[1-\frac{\omega^{2}}{\omega_{0}^{2}}]}{[1-\frac{\omega^{2}}{\omega_{0}^{2}}]^{\frac{1}{4}}}\tau)}(\frac{dz}{d\tau})^{2} \Big)d\tau^{2} - \nonumber\\&& \Big(e^{-2a\sigma}+ \frac{1}{\cosh^{2}(\frac{T_{0}ln^{-1}[1-\frac{\omega^{2}}{\omega_{0}^{2}}]}{[1-\frac{\omega^{2}}{\omega_{0}^{2}}]^{\frac{1}{4}}}\tau)}(\frac{dz}{d\sigma})^{2} \Big) d\sigma^{2} -\nonumber\\&& \frac{1}{\sinh(\frac{T_{0}ln^{-1}[1-\frac{\omega^{2}}{\omega_{0}^{2}}]}{[1-\frac{\omega^{2}}{\omega_{0}^{2}}]^{\frac{1}{4}}}\tau)\cosh(a\tau)}(\frac{dz}{d\tau }\frac{dz}{d\sigma})d\tau d\sigma + \nonumber\\&& \Big(\frac{1}{\frac{T_{0}ln^{-1}[1-\frac{\omega^{2}}{\omega_{0}^{2}}]}{[1-\frac{\omega^{2}}{\omega_{0}^{2}}]^{\frac{1}{4}}}}e^{-\frac{T_{0}ln^{-1}[1-\frac{\omega^{2}}{\omega_{0}^{2}}]}{[1-\frac{\omega^{2}}{\omega_{0}^{2}}]^{\frac{1}{4}}}\sigma} \cosh(\frac{T_{0}ln^{-1}[1-\frac{\omega^{2}}{\omega_{0}^{2}}]}{[1-\frac{\omega^{2}}{\omega_{0}^{2}}]^{\frac{1}{4}}}\tau)\times \nonumber\\&& \Big)^{2}\Big(d\theta^{2} + sin^{2}\theta d\phi^{2}\Big)  + \sum_{i=1}^{5}dx_{i}^{2} \nonumber\\&&
\label{thermal3}
\end{eqnarray}

Above metrics are corresponded to thermal rotating accretion disks. These metric depends on the temperature and rotating velocity of accretion disks.

To obtain the spectrum of rotating accretion disk, we should obtain the action. To this aim,we will use of concept of string model for accretion disk in \cite{w4}.  For flat space-time, the action of disk is \cite{w4,w5}:

\begin{eqnarray}
&& S_{3}=-T_{tri} \int d^{3}\sigma \sqrt{\eta^{ab}
	g_{MN}\partial_{a}X^{M}\partial_{b}X^{N}+2\pi
	l_{s}^{2}G(F))}\nonumber\\&&
G=(\sum_{n=1}^{3}\frac{1}{n!}(-\frac{F_{1}..F_{n}}{\beta^{2}}))
\nonumber\\&& F=F_{\mu\nu}F^{\mu\nu}\quad
F_{\mu\nu}=\partial_{\mu}A_{\nu}-
\partial_{\nu}A_{\mu}\label{f1}
\end{eqnarray}
where $g_{MN}$ is the background metric, $ X^{M}(\sigma^{a})$'s
are scalar fields , $\sigma^{a}$'s are
the accretion disk coordinates, $a, b = 0, 1, ..., 3$ are world-volume
indices of rotating accretion disk and $M,N=0, 1, ..., 10$ are eleven dimensional
spacetime indices. Also, $G$ is the nonlinear field
\cite{w5} and $A$ is the photon which exchanges between
disks. In a Rindler space-time, with the metrics in equations (\ref{a5},\ref{a6}), we obtain:

\begin{eqnarray}
&&  S_{I,end1} = -  \int dt \int_{\sigma_{0}}^{\infty}d\sigma \Big(\frac{1}{\frac{T_{0}ln^{-1}[1-\frac{\omega^{2}}{\omega_{0}^{2}}]}{[1-\frac{\omega^{2}}{\omega_{0}^{2}}]^{\frac{1}{4}}}
}e^{-\frac{T_{0}ln^{-1}[1-\frac{\omega^{2}}{\omega_{0}^{2}}]}{[1-\frac{\omega^{2}}{\omega_{0}^{2}}]^{\frac{1}{4}}}
\sigma}e^{\frac{T_{0}ln^{-1}[1-\frac{\omega^{2}}{\omega_{0}^{2}}]}{[1-\frac{\omega^{2}}{\omega_{0}^{2}}]^{\frac{1}{4}}}
\sigma} \cosh(\frac{T_{0}ln^{-1}[1-\frac{\omega^{2}}{\omega_{0}^{2}}]}{[1-\frac{\omega^{2}}{\omega_{0}^{2}}]^{\frac{1}{4}}}
\tau)\Big)^{2}\times \nonumber\\&&\Big(\sinh^{2}(\frac{T_{0}ln^{-1}[1-\frac{\omega^{2}}{\omega_{0}^{2}}]}{[1-\frac{\omega^{2}}{\omega_{0}^{2}}]^{\frac{1}{4}}}
\tau)+cosh^{2}(\frac{T_{0}ln^{-1}[1-\frac{\omega^{2}}{\omega_{0}^{2}}]}{[1-\frac{\omega^{2}}{\omega_{0}^{2}}]^{\frac{1}{4}}}
\tau)\Big)\times \nonumber\\&& [1 + \frac{e^{-2\frac{T_{0}ln^{-1}[1-\frac{\omega^{2}}{\omega_{0}^{2}}]}{[1-\frac{\omega^{2}}{\omega_{0}^{2}}]^{\frac{1}{4}}}
\sigma}}{\sinh^{2}(\frac{T_{0}ln^{-1}[1-\frac{\omega^{2}}{\omega_{0}^{2}}]}{[1-\frac{\omega^{2}}{\omega_{0}^{2}}]^{\frac{1}{4}}}
\tau)}(\frac{dz}{d\tau})^{2} + \frac{e^{-2\frac{T_{0}ln^{-1}[1-\frac{\omega^{2}}{\omega_{0}^{2}}]}{[1-\frac{\omega^{2}}{\omega_{0}^{2}}]^{\frac{1}{4}}}
\sigma}}{\cosh^{2}(\frac{T_{0}ln^{-1}[1-\frac{\omega^{2}}{\omega_{0}^{2}}]}{[1-\frac{\omega^{2}}{\omega_{0}^{2}}]^{\frac{1}{4}}}
\tau)}(\frac{dz}{d\sigma})^{2} +\nonumber\\&& \frac{e^{-2\frac{T_{0}ln^{-1}[1-\frac{\omega^{2}}{\omega_{0}^{2}}]}{[1-\frac{\omega^{2}}{\omega_{0}^{2}}]^{\frac{1}{4}}}
\sigma}}{\sinh(\frac{T_{0}ln^{-1}[1-\frac{\omega^{2}}{\omega_{0}^{2}}]}{[1-\frac{\omega^{2}}{\omega_{0}^{2}}]^{\frac{1}{4}}}
\tau)\cosh(\frac{T_{0}ln^{-1}[1-\frac{\omega^{2}}{\omega_{0}^{2}}]}{[1-\frac{\omega^{2}}{\omega_{0}^{2}}]^{\frac{1}{4}}}
\tau)}((\frac{dz}{d\tau }\frac{dz}{d\sigma})) - (2\pi l_{s}^{2}G(F)) ]^{1/2}\nonumber\\&&
\label{a10}
\end{eqnarray}

\begin{eqnarray}
&& S_{II,end1} = -  \int dt \int_{\sigma_{0}}^{\infty}d\sigma \Big(\frac{1}{\frac{T_{0}ln^{-1}[1-\frac{\omega^{2}}{\omega_{0}^{2}}]}{[1-\frac{\omega^{2}}{\omega_{0}^{2}}]^{\frac{1}{4}}}
}e^{-\frac{T_{0}ln^{-1}[1-\frac{\omega^{2}}{\omega_{0}^{2}}]}{[1-\frac{\omega^{2}}{\omega_{0}^{2}}]^{\frac{1}{4}}}
\sigma} \cosh(\frac{T_{0}ln^{-1}[1-\frac{\omega^{2}}{\omega_{0}^{2}}]}{[1-\frac{\omega^{2}}{\omega_{0}^{2}}]^{\frac{1}{4}}}
\tau)\Big)^{2}\times \nonumber\\&&\Big(\sinh^{2}(\frac{T_{0}ln^{-1}[1-\frac{\omega^{2}}{\omega_{0}^{2}}]}{[1-\frac{\omega^{2}}{\omega_{0}^{2}}]^{\frac{1}{4}}}
\tau)+cosh^{2}(\frac{T_{0}ln^{-1}[1-\frac{\omega^{2}}{\omega_{0}^{2}}]}{[1-\frac{\omega^{2}}{\omega_{0}^{2}}]^{\frac{1}{4}}}
\tau)\Big)\times \nonumber\\&& [1 + \frac{e^{2\frac{T_{0}ln^{-1}[1-\frac{\omega^{2}}{\omega_{0}^{2}}]}{[1-\frac{\omega^{2}}{\omega_{0}^{2}}]^{\frac{1}{4}}}
\sigma}}{\sinh^{2}(\frac{T_{0}ln^{-1}[1-\frac{\omega^{2}}{\omega_{0}^{2}}]}{[1-\frac{\omega^{2}}{\omega_{0}^{2}}]^{\frac{1}{4}}}
\tau)}(\frac{dz}{d\tau})^{2} + \frac{e^{2a\sigma}}{\cosh^{2}(\frac{T_{0}ln^{-1}[1-\frac{\omega^{2}}{\omega_{0}^{2}}]}{[1-\frac{\omega^{2}}{\omega_{0}^{2}}]^{\frac{1}{4}}}
\tau)}(\frac{dz}{d\sigma})^{2} - \nonumber\\&& \frac{e^{2a\sigma}}{\sinh(\frac{T_{0}ln^{-1}[1-\frac{\omega^{2}}{\omega_{0}^{2}}]}{[1-\frac{\omega^{2}}{\omega_{0}^{2}}]^{\frac{1}{4}}}
\tau)\cosh(\frac{T_{0}ln^{-1}[1-\frac{\omega^{2}}{\omega_{0}^{2}}]}{[1-\frac{\omega^{2}}{\omega_{0}^{2}}]^{\frac{1}{4}}}
\tau)}(\frac{dz}{d\tau }\frac{dz}{d\sigma}) - (2\pi l_{s}^{2}G(F)) ] \nonumber\\&&
\label{a11}
\end{eqnarray}

 Using the method in ref \cite{w4}, we can obtain the energy for accretion disks:

\begin{eqnarray}
&& E_{I,end1}=  \int d^{3}\sigma \varrho_{I,end1} \nonumber\\&& \varrho_{I,end1} = \int d^{3}\sigma  [1 + \frac{e^{-2\frac{T_{0}ln^{-1}[1-\frac{\omega^{2}}{\omega_{0}^{2}}]}{[1-\frac{\omega^{2}}{\omega_{0}^{2}}]^{\frac{1}{4}}}
\sigma}}{\sinh^{2}(\frac{T_{0}ln^{-1}[1-\frac{\omega^{2}}{\omega_{0}^{2}}]}{[1-\frac{\omega^{2}}{\omega_{0}^{2}}]^{\frac{1}{4}}}
\tau)}(\frac{dz}{d\tau})^{2} + \frac{e^{-2\frac{T_{0}ln^{-1}[1-\frac{\omega^{2}}{\omega_{0}^{2}}]}{[1-\frac{\omega^{2}}{\omega_{0}^{2}}]^{\frac{1}{4}}}
\sigma}}{\cosh^{2}(\frac{T_{0}ln^{-1}[1-\frac{\omega^{2}}{\omega_{0}^{2}}]}{[1-\frac{\omega^{2}}{\omega_{0}^{2}}]^{\frac{1}{4}}}
\tau)}(\frac{dz}{d\sigma})^{2} + \nonumber\\&& \frac{e^{-2\frac{T_{0}ln^{-1}[1-\frac{\omega^{2}}{\omega_{0}^{2}}]}{[1-\frac{\omega^{2}}{\omega_{0}^{2}}]^{\frac{1}{4}}}
\sigma}}{\sinh(\frac{T_{0}ln^{-1}[1-\frac{\omega^{2}}{\omega_{0}^{2}}]}{[1-\frac{\omega^{2}}{\omega_{0}^{2}}]^{\frac{1}{4}}}
\tau)\cosh(\frac{T_{0}ln^{-1}[1-\frac{\omega^{2}}{\omega_{0}^{2}}]}{[1-\frac{\omega^{2}}{\omega_{0}^{2}}]^{\frac{1}{4}}}
\tau)}((\frac{dz}{d\tau }\frac{dz}{d\sigma}))]^{1/2} O_{tot,I}  \nonumber\\ &&
O_{tot,I} =[1+\frac{k^{2}_{3}}{ \Big(\frac{1}{\frac{T_{0}ln^{-1}[1-\frac{\omega^{2}}{\omega_{0}^{2}}]}{[1-\frac{\omega^{2}}{\omega_{0}^{2}}]^{\frac{1}{4}}}
}e^{\frac{T_{0}ln^{-1}[1-\frac{\omega^{2}}{\omega_{0}^{2}}]}{[1-\frac{\omega^{2}}{\omega_{0}^{2}}]^{\frac{1}{4}}}
\sigma} \cosh(\frac{T_{0}ln^{-1}[1-\frac{\omega^{2}}{\omega_{0}^{2}}]}{[1-\frac{\omega^{2}}{\omega_{0}^{2}}]^{\frac{1}{4}}}
\tau)\Big)^{4}}]^{1/2}\times \nonumber\\&&[1+\frac{k^{2}_{2}}{ \Big(\frac{1}{\frac{T_{0}ln^{-1}[1-\frac{\omega^{2}}{\omega_{0}^{2}}]}{[1-\frac{\omega^{2}}{\omega_{0}^{2}}]^{\frac{1}{4}}}
}e^{\frac{T_{0}ln^{-1}[1-\frac{\omega^{2}}{\omega_{0}^{2}}]}{[1-\frac{\omega^{2}}{\omega_{0}^{2}}]^{\frac{1}{4}}}
\sigma} \cosh(\frac{T_{0}ln^{-1}[1-\frac{\omega^{2}}{\omega_{0}^{2}}]}{[1-\frac{\omega^{2}}{\omega_{0}^{2}}]^{\frac{1}{4}}}
\tau)\Big)^{4}}]^{1/2}
\label{a14}
\end{eqnarray}

\begin{eqnarray}
&& E_{II,end2}= \int d^{3}\sigma \varrho_{II,end2} \nonumber\\&& \varrho_{II,end2} =\nonumber\\&&\int d^{3}\sigma [1 + \frac{e^{2\frac{T_{0}ln^{-1}[1-\frac{\omega^{2}}{\omega_{0}^{2}}]}{[1-\frac{\omega^{2}}{\omega_{0}^{2}}]^{\frac{1}{4}}}
\sigma}}{\sinh^{2}(\frac{T_{0}ln^{-1}[1-\frac{\omega^{2}}{\omega_{0}^{2}}]}{[1-\frac{\omega^{2}}{\omega_{0}^{2}}]^{\frac{1}{4}}}
\tau)}(\frac{dz}{d\tau})^{2} +\nonumber\\&& \frac{e^{2\frac{T_{0}ln^{-1}[1-\frac{\omega^{2}}{\omega_{0}^{2}}]}{[1-\frac{\omega^{2}}{\omega_{0}^{2}}]^{\frac{1}{4}}}
\sigma}}{\cosh^{2}(\frac{T_{0}ln^{-1}[1-\frac{\omega^{2}}{\omega_{0}^{2}}]}{[1-\frac{\omega^{2}}{\omega_{0}^{2}}]^{\frac{1}{4}}}
\tau)}(\frac{dz}{d\sigma})^{2} - \frac{e^{2\frac{T_{0}ln^{-1}[1-\frac{\omega^{2}}{\omega_{0}^{2}}]}{[1-\frac{\omega^{2}}{\omega_{0}^{2}}]^{\frac{1}{4}}}
\sigma}}{\sinh(\frac{T_{0}ln^{-1}[1-\frac{\omega^{2}}{\omega_{0}^{2}}]}{[1-\frac{\omega^{2}}{\omega_{0}^{2}}]^{\frac{1}{4}}}
\tau)\cosh(\frac{T_{0}ln^{-1}[1-\frac{\omega^{2}}{\omega_{0}^{2}}]}{[1-\frac{\omega^{2}}{\omega_{0}^{2}}]^{\frac{1}{4}}}
\tau)}((\frac{dz}{d\tau }\frac{dz}{d\sigma}))]^{1/2} O_{tot,II} \nonumber\\ &&O_{tot,II} =[1+\frac{k^{2}_{3}}{ \Big(\frac{1}{\frac{T_{0}ln^{-1}[1-\frac{\omega^{2}}{\omega_{0}^{2}}]}{[1-\frac{\omega^{2}}{\omega_{0}^{2}}]^{\frac{1}{4}}}
}e^{-\frac{T_{0}ln^{-1}[1-\frac{\omega^{2}}{\omega_{0}^{2}}]}{[1-\frac{\omega^{2}}{\omega_{0}^{2}}]^{\frac{1}{4}}}
\sigma} \cosh(\frac{T_{0}ln^{-1}[1-\frac{\omega^{2}}{\omega_{0}^{2}}]}{[1-\frac{\omega^{2}}{\omega_{0}^{2}}]^{\frac{1}{4}}}
\tau)\Big)^{4}}]^{1/2}\times \nonumber\\&&\sqrt{1+\frac{k^{2}_{2}}{\Big(\frac{1}{\frac{T_{0}ln^{-1}[1-\frac{\omega^{2}}{\omega_{0}^{2}}]}{[1-\frac{\omega^{2}}{\omega_{0}^{2}}]^{\frac{1}{4}}}
}e^{-\frac{T_{0}ln^{-1}[1-\frac{\omega^{2}}{\omega_{0}^{2}}]}{[1-\frac{\omega^{2}}{\omega_{0}^{2}}]^{\frac{1}{4}}}
\sigma} \cosh(\frac{T_{0}ln^{-1}[1-\frac{\omega^{2}}{\omega_{0}^{2}}]}{[1-\frac{\omega^{2}}{\omega_{0}^{2}}]^{\frac{1}{4}}}
\tau)\Big)^{4}}}
\label{a15}
\end{eqnarray}

Above energies depend on the rotating velocities of disks. In a rotating BIon, when, energy of one disk increases, energy of other disk deceases. We can obtain waves equations from above equations:

\begin{eqnarray}
&&\frac{\partial \varrho_{I,end1}}{\partial ( \frac{\partial z }{\partial \tau} )}- \frac{\partial \varrho_{I,end1}}{\partial z } =0\\&& \frac{\partial \varrho_{II,end1}}{\partial ( \frac{\partial z }{\partial \tau} )}- \frac{\partial \varrho_{II,end1}}{\partial z } =0
\label{eq1}
\end{eqnarray}

Solving above equation, we obtain:

\begin{eqnarray}
&& z_{I-A,end1} =z_{II-B,end1}\simeq\int d\tau d\sigma \Big(e^{-4 \frac{T_{0}ln^{-1}[1-\frac{\omega^{2}}{\omega_{0}^{2}}]}{[1-\frac{\omega^{2}}{\omega_{0}^{2}}]^{\frac{1}{4}}}\sigma}\sinh^{2}( \frac{T_{0}ln^{-1}[1-\frac{\omega^{2}}{\omega_{0}^{2}}]}{[1-\frac{\omega^{2}}{\omega_{0}^{2}}]^{\frac{1}{4}}}\tau)\cosh^{2}( \frac{T_{0}ln^{-1}[1-\frac{\omega^{2}}{\omega_{0}^{2}}]}{[1-\frac{\omega^{2}}{\omega_{0}^{2}}]^{\frac{1}{4}}}\tau)\Big) \times \nonumber\\&& \Big(\frac{O_{tot,I}(\tau,\sigma)\Big(\frac{O_{tot,I}(\tau,\sigma)}{O_{tot,I}(\tau,\sigma_{0})}-e^{-4 \frac{T_{0}ln^{-1}[1-\frac{\omega^{2}}{\omega_{0}^{2}}]}{[1-\frac{\omega^{2}}{\omega_{0}^{2}}]^{\frac{1}{4}}}(\sigma-\sigma_{0})}\frac{\cosh^{2}( \frac{T_{0}ln^{-1}[1-\frac{\omega^{2}}{\omega_{0}^{2}}]}{[1-\frac{\omega^{2}}{\omega_{0}^{2}}]^{\frac{1}{4}}}\tau_{0})}{\cosh^{2}( \frac{T_{0}ln^{-1}[1-\frac{\omega^{2}}{\omega_{0}^{2}}]}{[1-\frac{\omega^{2}}{\omega_{0}^{2}}]^{\frac{1}{4}}}\tau)}\Big)^{-\frac{1}{2}}}{O_{tot,I}(\tau_{0},\sigma)\Big(\frac{O_{tot,I}(\tau_{0},\sigma)}{O_{tot,I}(\tau_{0},\sigma_{0})}-e^{-4a(\sigma-\sigma_{0})}\frac{\cosh^{2}(a\tau_{0})}{\cosh^{2}( \frac{T_{0}ln^{-1}[1-\frac{\omega^{2}}{\omega_{0}^{2}}]}{[1-\frac{\omega^{2}}{\omega_{0}^{2}}]^{\frac{1}{4}}}\tau)}\Big)^{-\frac{1}{2}}}-\frac{\sinh^{2}(a\tau_{0})}{\sinh^{2}( \frac{T_{0}ln^{-1}[1-\frac{\omega^{2}}{\omega_{0}^{2}}]}{[1-\frac{\omega^{2}}{\omega_{0}^{2}}]^{\frac{1}{4}}}\tau)}\Big)^{-\frac{1}{2}}
\label{eqa19}
\end{eqnarray}

\begin{eqnarray}
&& z_{I-B,end1} =z_{II-A,end1}\simeq\int d\tau d\sigma \Big(e^{4 \frac{T_{0}ln^{-1}[1-\frac{\omega^{2}}{\omega_{0}^{2}}]}{[1-\frac{\omega^{2}}{\omega_{0}^{2}}]^{\frac{1}{4}}}\sigma}\sinh^{2}( \frac{T_{0}ln^{-1}[1-\frac{\omega^{2}}{\omega_{0}^{2}}]}{[1-\frac{\omega^{2}}{\omega_{0}^{2}}]^{\frac{1}{4}}}\tau)\cosh^{2}( \frac{T_{0}ln^{-1}[1-\frac{\omega^{2}}{\omega_{0}^{2}}]}{[1-\frac{\omega^{2}}{\omega_{0}^{2}}]^{\frac{1}{4}}}\tau)\Big) \times \nonumber\\&& \Big(\frac{O_{tot,II}(\tau,\sigma)\Big(\frac{O_{tot,II}(\tau,\sigma)}{O_{tot,II}(\tau,\sigma_{0})}-e^{4 \frac{T_{0}ln^{-1}[1-\frac{\omega^{2}}{\omega_{0}^{2}}]}{[1-\frac{\omega^{2}}{\omega_{0}^{2}}]^{\frac{1}{4}}}(\sigma-\sigma_{0})}\frac{\cosh^{2}( \frac{T_{0}ln^{-1}[1-\frac{\omega^{2}}{\omega_{0}^{2}}]}{[1-\frac{\omega^{2}}{\omega_{0}^{2}}]^{\frac{1}{4}}}\tau_{0})}{\cosh^{2}( \frac{T_{0}ln^{-1}[1-\frac{\omega^{2}}{\omega_{0}^{2}}]}{[1-\frac{\omega^{2}}{\omega_{0}^{2}}]^{\frac{1}{4}}}\tau)}\Big)^{-\frac{1}{2}}}{O_{tot,II}(\tau_{0},\sigma)\Big(\frac{F_{DBI,II,A}(\tau_{0},\sigma)}{O_{tot,II}(\tau_{0},\sigma_{0})}-e^{4 \frac{T_{0}ln^{-1}[1-\frac{\omega^{2}}{\omega_{0}^{2}}]}{[1-\frac{\omega^{2}}{\omega_{0}^{2}}]^{\frac{1}{4}}}(\sigma-\sigma_{0})}\frac{\cosh^{2}( \frac{T_{0}ln^{-1}[1-\frac{\omega^{2}}{\omega_{0}^{2}}]}{[1-\frac{\omega^{2}}{\omega_{0}^{2}}]^{\frac{1}{4}}}\tau_{0})}{\cosh^{2}( \frac{T_{0}ln^{-1}[1-\frac{\omega^{2}}{\omega_{0}^{2}}]}{[1-\frac{\omega^{2}}{\omega_{0}^{2}}]^{\frac{1}{4}}}\tau)}\Big)^{-\frac{1}{2}}}-\frac{\sinh^{2}(a\tau_{0})}{\sinh^{2}( \frac{T_{0}ln^{-1}[1-\frac{\omega^{2}}{\omega_{0}^{2}}]}{[1-\frac{\omega^{2}}{\omega_{0}^{2}}]^{\frac{1}{4}}}\tau)}\Big)^{-\frac{1}{2}}
\label{eqa20}
\end{eqnarray}

Above results show that length of accretion disks depend on temperature and rotating velocity. Using equation (\ref{a2}), we can also obtain  the radius of disks:

\begin{eqnarray}
&& t_{I,end1}= \frac{1}{ \frac{T_{0}ln^{-1}[1-\frac{\omega^{2}}{\omega_{0}^{2}}]}{[1-\frac{\omega^{2}}{\omega_{0}^{2}}]^{\frac{1}{4}}}}e^{ \frac{T_{0}ln^{-1}[1-\frac{\omega^{2}}{\omega_{0}^{2}}]}{[1-\frac{\omega^{2}}{\omega_{0}^{2}}]^{\frac{1}{4}}}\sigma} \sinh( \frac{T_{0}ln^{-1}[1-\frac{\omega^{2}}{\omega_{0}^{2}}]}{[1-\frac{\omega^{2}}{\omega_{0}^{2}}]^{\frac{1}{4}}}\tau) \nonumber\\&& r_{I,end1}=\frac{1}{ \frac{T_{0}ln^{-1}[1-\frac{\omega^{2}}{\omega_{0}^{2}}]}{[1-\frac{\omega^{2}}{\omega_{0}^{2}}]^{\frac{1}{4}}}}e^{ \frac{T_{0}ln^{-1}[1-\frac{\omega^{2}}{\omega_{0}^{2}}]}{[1-\frac{\omega^{2}}{\omega_{0}^{2}}]^{\frac{1}{4}}}\sigma} \cosh( \frac{T_{0}ln^{-1}[1-\frac{\omega^{2}}{\omega_{0}^{2}}]}{[1-\frac{\omega^{2}}{\omega_{0}^{2}}]^{\frac{1}{4}}}\tau) \quad \text{In Region I} \nonumber\\&& t_{II,end1}= - \frac{1}{ \frac{T_{0}ln^{-1}[1-\frac{\omega^{2}}{\omega_{0}^{2}}]}{[1-\frac{\omega^{2}}{\omega_{0}^{2}}]^{\frac{1}{4}}}}e^{- \frac{T_{0}ln^{-1}[1-\frac{\omega^{2}}{\omega_{0}^{2}}]}{[1-\frac{\omega^{2}}{\omega_{0}^{2}}]^{\frac{1}{4}}}\sigma} \sinh( \frac{T_{0}ln^{-1}[1-\frac{\omega^{2}}{\omega_{0}^{2}}]}{[1-\frac{\omega^{2}}{\omega_{0}^{2}}]^{\frac{1}{4}}}\tau) \nonumber\\&& r_{II,end1} = \frac{1}{ \frac{T_{0}ln^{-1}[1-\frac{\omega^{2}}{\omega_{0}^{2}}]}{[1-\frac{\omega^{2}}{\omega_{0}^{2}}]^{\frac{1}{4}}}} e^{- \frac{T_{0}ln^{-1}[1-\frac{\omega^{2}}{\omega_{0}^{2}}]}{[1-\frac{\omega^{2}}{\omega_{0}^{2}}]^{\frac{1}{4}}}\sigma} \cosh( \frac{T_{0}ln^{-1}[1-\frac{\omega^{2}}{\omega_{0}^{2}}]}{[1-\frac{\omega^{2}}{\omega_{0}^{2}}]^{\frac{1}{4}}}\tau) \quad \text{In Region II}
\label{ta2}
\end{eqnarray}

Using above coordinates in equations (\ref{eqa19}, \ref{eqa20},\ref{ta2})

\begin{eqnarray}
&& A_{I,end1} = \pi r_{I,end1}^{2}(z_{I-A,end1} +z_{I-B,end1}) \nonumber\\&& A_{II,end1} = \pi r_{II,end1}^{2}(z_{II-A,end1} +z_{II-B,end1})
\label{tBa2}
\end{eqnarray}

Above results show dependency of accretion disk to temperature and rotating velocity. Area in region I and II are equal. Because, length A in region I acts like the length B in region II and also, length B in region I acts like the length A in region II. By increasing  rotating velocity, area of disks increase (See figure 2).

\section{Contraction of one accreion disk of a rotating BIon in a Rindler space-time }\label{o2}

In this section,we consider contacting disks. For these types of disks, the relation between acceleration and temperature is \cite{w4,q12}:

\begin{figure*}[thbp]
	\begin{center}
		\begin{tabular}{rl}
			\includegraphics[width=14cm]{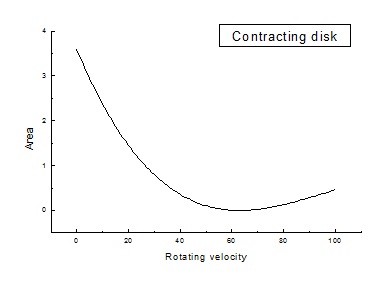}
		\end{tabular}
	\end{center}
	\caption{Contracting disk. Area in terms of rotating velocity}
\end{figure*}

\begin{eqnarray}
&& a \sim 2\pi T= 2\pi T_{0}ln[1-\frac{\omega^{2}}{\omega_{0}^{2}}]
\label{SEC1}
\end{eqnarray}

where $T_{0}$ is temperature of non-rotating accretion disk and $\omega$ is the rotating velocity. Following methods in previous section, we obtain energy for contracting disks.

\begin{eqnarray}
&& E_{I,end2}= \int d^{3}\sigma \varrho_{I,end2} \nonumber\\&& \varrho_{I,end2} = \int d^{3}\sigma \times\\&& \Big(1 + \frac{e^{- 2\pi T_{0}\sqrt{1-\frac{l_{0}^{2}\omega^{2}}{c^{2}}\sigma}}}{\sinh^{2}(\frac{T_{0}ln^{-1}[1-\frac{\omega^{2}}{\omega_{0}^{2}}]}{[1-\frac{\omega^{2}}{\omega_{0}^{2}}]^{\frac{1}{4}}}
\tau)}(\frac{dz}{d\tau})^{2} +  \frac{e^{- 2\pi T_{0}ln[1-\frac{\omega^{2}}{\omega_{0}^{2}}]
\sigma}}{\cosh^{2}( 2\pi T_{0}ln[1-\frac{\omega^{2}}{\omega_{0}^{2}}]\tau)}(\frac{dz}{d\sigma})^{2} + \nonumber\\&& \frac{e^{- 2\pi T_{0}ln[1-\frac{\omega^{2}}{\omega_{0}^{2}}]\sigma}}{\sinh( 2\pi T_{0}ln[1-\frac{\omega^{2}}{\omega_{0}^{2}}]\tau)\cosh( 2\pi T_{0}ln[1-\frac{\omega^{2}}{\omega_{0}^{2}}]
\tau)}((\frac{dz}{d\tau }\frac{dz}{d\sigma}))\Big) O_{tot,I,end2}  \nonumber\\ &&
O_{tot,I,end2} =\sqrt{1+\frac{k^{2}_{3}}{ \Big(\frac{1}{ 2\pi T_{0}ln[1-\frac{\omega^{2}}{\omega_{0}^{2}}]\sigma} \cosh( 2\pi T_{0}ln[1-\frac{\omega^{2}}{\omega_{0}^{2}}]
\tau)\Big)^{4}}}\times \nonumber\\&&\sqrt{1+\frac{k^{2}_{2}}{ \Big(\frac{1}{ 2\pi T_{0}ln[1-\frac{\omega^{2}}{\omega_{0}^{2}}]
\sigma} \cosh( 2\pi T_{0}ln[1-\frac{\omega^{2}}{\omega_{0}^{2}}]
\tau)\Big)^{4}}}
\label{SECa14}
\end{eqnarray}

\begin{eqnarray}
&& E_{II,end2}= \int d^{3}\sigma \varrho_{II,end2} \nonumber\\&& \varrho_{II,end2} = \int d^{3}\sigma \times\\&& \Big(1 + \frac{e^{ 2\pi T_{0}\sqrt{1-\frac{l_{0}^{2}\omega^{2}}{c^{2}}\sigma}}}{\sinh^{2}(\frac{T_{0}ln^{-1}[1-\frac{\omega^{2}}{\omega_{0}^{2}}]}{[1-\frac{\omega^{2}}{\omega_{0}^{2}}]^{\frac{1}{4}}}
\tau)}(\frac{dz}{d\tau})^{2} +  \frac{e^{ 2\pi T_{0}ln[1-\frac{\omega^{2}}{\omega_{0}^{2}}]
\sigma}}{\cosh^{2}( 2\pi T_{0}ln[1-\frac{\omega^{2}}{\omega_{0}^{2}}]\tau)}(\frac{dz}{d\sigma})^{2} -\nonumber\\&& \frac{e^{-2\pi T_{0}ln[1-\frac{\omega^{2}}{\omega_{0}^{2}}]\sigma}}{\sinh( 2\pi T_{0}ln[1-\frac{\omega^{2}}{\omega_{0}^{2}}]\tau)\cosh( 2\pi T_{0}ln[1-\frac{\omega^{2}}{\omega_{0}^{2}}]
\tau)}((\frac{dz}{d\tau }\frac{dz}{d\sigma}))\Big) O_{tot,II,end2}  \nonumber\\ &&
O_{tot,II} =\sqrt{1+\frac{k^{2}_{3}}{ \Big(\frac{1}{ -2\pi T_{0}ln[1-\frac{\omega^{2}}{\omega_{0}^{2}}]\sigma} \cosh( -2\pi T_{0}ln[1-\frac{\omega^{2}}{\omega_{0}^{2}}]
\tau)\Big)^{4}}}\times \nonumber\\&&\sqrt{1+\frac{k^{2}_{2}}{ \Big(\frac{1}{- 2\pi T_{0}ln[1-\frac{\omega^{2}}{\omega_{0}^{2}}]
\sigma} \cosh( -2\pi T_{0}ln[1-\frac{\omega^{2}}{\omega_{0}^{2}}]
\tau)\Big)^{4}}}
\label{SECa15}
\end{eqnarray}

Above two energies act reverse to each other.When energy of one end increases, energy of other end decreases. We can obtain wave equation from above equations\cite{w4}:

\begin{eqnarray}
&&\frac{\partial \varrho_{I,end2}}{\partial ( \frac{\partial z }{\partial \tau} )}- \frac{\partial \varrho_{I,end2}}{\partial z } =0\\&& \frac{\partial \varrho_{II,end2}}{\partial ( \frac{\partial z }{\partial \tau} )}- \frac{\partial \varrho_{II,end2}}{\partial z } =0
\label{eq2}
\end{eqnarray}

Solving above equation, we obtain lengths of disks:

\begin{eqnarray}
&& z_{I-A,end2} =z_{II-B,end2}\simeq\int d\tau d\sigma \Big(e^{-4 2\pi T_{0}ln[1-\frac{\omega^{2}}{\omega_{0}^{2}}]\sigma}\sinh^{2}( 2\pi T_{0}ln[1-\frac{\omega^{2}}{\omega_{0}^{2}}]\tau)\cosh^{2}( 2\pi T_{0}ln[1-\frac{\omega^{2}}{\omega_{0}^{2}}]\tau)\Big) \times \nonumber\\&& \Big(\frac{O_{tot,I,end2} (\tau,\sigma)\Big(\frac{O_{tot,I,end2} (\tau,\sigma)}{O_{tot,I,end2} (\tau,\sigma_{0})}-e^{-4 2\pi T_{0}ln[1-\frac{\omega^{2}}{\omega_{0}^{2}}](\sigma-\sigma_{0})}\frac{\cosh^{2}( 2\pi T_{0}ln[1-\frac{\omega^{2}}{\omega_{0}^{2}}]\tau_{0})}{\cosh^{2}( 2\pi T_{0}ln[1-\frac{\omega^{2}}{\omega_{0}^{2}}]\tau)}\Big)^{-\frac{1}{2}}}{O_{tot,I,end2} (\tau_{0},\sigma)\Big(\frac{O_{tot,I,end2} (\tau_{0},\sigma)}{O_{tot,I,end2} (\tau_{0},\sigma_{0})}-e^{-4 2\pi T_{0}ln[1-\frac{\omega^{2}}{\omega_{0}^{2}}](\sigma-\sigma_{0})}\frac{\cosh^{2}(a\tau_{0})}{\cosh^{2}( 2\pi T_{0}ln[1-\frac{\omega^{2}}{\omega_{0}^{2}}]\tau)}\Big)^{-\frac{1}{2}}}-\frac{\sinh^{2}( 2\pi T_{0}ln[1-\frac{\omega^{2}}{\omega_{0}^{2}}]\tau_{0})}{\sinh^{2}( 2\pi T_{0}ln[1-\frac{\omega^{2}}{\omega_{0}^{2}}]\tau)}\Big)^{-\frac{1}{2}}
\label{ta19}
\end{eqnarray}

\begin{eqnarray}
&& z_{I-B,end2} =z_{II-A,end2}\simeq\int d\tau d\sigma \Big(e^{4 2\pi T_{0}ln[1-\frac{\omega^{2}}{\omega_{0}^{2}}]\sigma}\sinh^{2}( 2\pi T_{0}ln[1-\frac{\omega^{2}}{\omega_{0}^{2}}]\tau)\cosh^{2}( 2\pi T_{0}ln[1-\frac{\omega^{2}}{\omega_{0}^{2}}]\tau)\Big) \times \nonumber\\&& \Big(\frac{O_{tot,II,end2}(\tau,\sigma)\Big(\frac{O_{tot,II,end2}(\tau,\sigma)}{O_{tot,II,end2}(\tau,\sigma_{0})}-e^{4 2\pi T_{0}ln[1-\frac{\omega^{2}}{\omega_{0}^{2}}](\sigma-\sigma_{0})}\frac{\cosh^{2}( 2\pi T_{0}ln[1-\frac{\omega^{2}}{\omega_{0}^{2}}]\tau_{0})}{\cosh^{2}( 2\pi T_{0}ln[1-\frac{\omega^{2}}{\omega_{0}^{2}}]\tau)}\Big)^{-\frac{1}{2}}}{O_{tot,II,end2}(\tau_{0},\sigma)\Big(\frac{O_{tot,II,end2}(\tau_{0},\sigma)}{O_{tot,II,end2}(\tau_{0},\sigma_{0})}-e^{4a(\sigma-\sigma_{0})}\frac{\cosh^{2} 2\pi T_{0}ln[1-\frac{\omega^{2}}{\omega_{0}^{2}}]a\tau_{0})}{\cosh^{2}( 2\pi T_{0}ln[1-\frac{\omega^{2}}{\omega_{0}^{2}}]\tau)}\Big)^{-\frac{1}{2}}}-\frac{\sinh^{2}( 2\pi T_{0}ln[1-\frac{\omega^{2}}{\omega_{0}^{2}}]\tau_{0})}{\sinh^{2}( 2\pi T_{0}ln[1-\frac{\omega^{2}}{\omega_{0}^{2}}]\tau)}\Big)^{-\frac{1}{2}}
\label{ta20}
\end{eqnarray}

Also, similar to previous section, we obtain relation between time and radius of disk as below:
\begin{eqnarray}
&& t_{I,end2}=\frac{1}{2\pi T_{0}ln[1-\frac{\omega^{2}}{\omega_{0}^{2}}]} e^{2\pi T_{0}ln[1-\frac{\omega^{2}}{\omega_{0}^{2}}]\sigma} \sinh(2\pi T_{0}ln[1-\frac{\omega^{2}}{\omega_{0}^{2}}]\tau) \nonumber\\&& r_{I,end2}=\frac{1}{2\pi T_{0}ln[1-\frac{\omega^{2}}{\omega_{0}^{2}}]}e^{2\pi T_{0}ln[1-\frac{\omega^{2}}{\omega_{0}^{2}}]\sigma} \cosh(2\pi T_{0}ln[1-\frac{\omega^{2}}{\omega_{0}^{2}}]\tau) \quad \nonumber\\&& \quad\text{In Region I} \nonumber\\&& t_{II,end2}= \frac{1}{2\pi T_{0}ln[1-\frac{\omega^{2}}{\omega_{0}^{2}}]}e^{-2\pi T_{0}ln[1-\frac{\omega^{2}}{\omega_{0}^{2}}]\sigma} \sinh(2\pi T_{0}ln[1-\frac{\omega^{2}}{\omega_{0}^{2}}]\tau) \nonumber\\&& r_{II,end2} = \frac{1}{2\pi T_{0}ln[1-\frac{\omega^{2}}{\omega_{0}^{2}}]} e^{-2\pi T_{0}ln[1-\frac{\omega^{2}}{\omega_{0}^{2}}]\sigma} \cosh(2\pi T_{0}ln[1-\frac{\omega^{2}}{\omega_{0}^{2}}]\tau) \nonumber\\&& \quad \text{In Region II}
\label{Ca2}
\end{eqnarray}

Using above coordinates, we can obtain area of contracting disks:

\begin{eqnarray}
&& A_{I,end2} = \pi r_{I,end2}^{2}(z_{I-A,end2} +z_{I-B,end2}) \nonumber\\&& A_{II,end2} = \pi r_{II,end2}^{2}(z_{II-A,end2} +z_{II-B,end2})
\label{tBa2}
\end{eqnarray}

Above results indicate dependency of contracting accretion disk to temperature and rotating velocity. Again, areas in region I and II are equal. Because, length A in region I acts like the length B in region II and also, length B in region I acts like the length A in region II. By increasing  rotating velocity, area of disks decreases (See figure 3).

\section{Summary} \label{sum}

In this research, we have proposed a mathematical model for  accretion disks which are emerged in a Rindler space-time. We have shown that each accretion disk connects to another disk and forms a thermal BIon. In a rotating BIon in each region, two accretion disks act reverse to each other. This means that with expansion of one disk, another one contracts. Also, accretion disk of region I of Rindler space-time acts revere to similar accretion disk in region II.
\section*{Acknowledgements}
\noindent Authors have no conflict of interest. Contributions of all authors are the same. The work of A.Sepehri has been supported financiall by research institute for Astronomy-Astrophysics of Maragha (RIAAM) under research project number 1/5750-20.

\end{document}